\documentclass[numberedheadings]{aip-cp}

\usepackage[numbers,elide]{natbib}
\usepackage{fancybox}
\usepackage{shortvrb}
\MakeShortVerb\|

\usepackage{listings}
\usepackage[usenames]{color}
\definecolor{codecolor}{gray}{.9}
\definecolor{rlcolor}{cmyk}{0,1,0,0}
\usepackage{enumerate}

\makeatletter
\def\@@tablenote[#1]#2{%
  \g@addto@macro\PrintTableNotes{\tabnotefont\leftskip6pt%
    \refstepcounter{tabnotecount}%
    \protected@xdef\@currentlabel{\thetabnotecount}%
    \label{#1}%
    {\textsuperscript{\thetabnotecount\hskip4pt}}#2\par%
  }%
}
\makeatother

\begin{document}

\title{Simulating Electromagnetic Cascades\\in Magnetospheres of Active Galactic Nuclei}

\author[aff1]{Christoph Wendel\corref{cor1}}
\author[aff1]{Dorit Glawion}
\author[aff1]{Amit Shukla}
\author[aff1]{Karl Mannheim}
\affil[aff1]{Universit\"at W\"urzburg, D-97074 W\"urzburg, Germany}
\corresp[cor1]{Corresponding author: cwendel@astro.uni-wuerzburg.de}

\maketitle

\begin{abstract}\\
\textsl{Context}: At the low accretion-rates typical for BL Lac-objects, magnetospheres of active galactic nuclei can develop vacuum gaps with strong electric fields accelerating charged seed particles parallel to the magnetic fields up to ultra-relativistic energies. The seed particles sustain electromagnetic cascades by inverse-Compton-scattering and subsequent pair-production in soft background-radiation-fields from the accretion-disk and/or photo-ionised clouds, along the direction of the primary particle beams.\\
\textsl{Method}: The one-dimensional kinetic equation describing this linear inverse-Compton-Klein-Nishina-pair-cascade is inferred. We have developed a novel code, that can numerically solve this kinetic equation for an ample variety of input-parameters. By this, quasi-stationary particle- and photon-spectra are obtained.\\
\textsl{Application}: We use the code to model the cascaded interaction of electrons, that have been accelerated in a vacuum gap in the magnetosphere of Mrk 501, with Lyman-alpha-photons. The resulting spectrum on top of a synchrotron-self-Compton-background can cause a narrow TeV-bump in the spectrum of Mrk 501.
\end{abstract}

\section{INTRODUCTION}

In astrophysics there are numerous scenarios (cf. \citet{Zdz}, henceforth Z88) in which relativistic electrons and/or high-energetic photons (HEPs) interact on soft background-photons (SBPs). Electrons inverse-Compton- (IC-) up-scatter the SBPs, thus producing HEPs. Simultaneously, the HEPs react with the SBPs and pair-produce new electron-positron-pairs. Obviously, both processes sustain each other, thus giving rise to the development of an electromagnetic cascade.\\
Especially, electrons that cause an electromagnetic cascade, can be abundant at the base of a jet of an active galactic nucleus (AGN). In the magnetosphere's poles, charge-depleted regions (called vacuum gaps or electrostatic gaps or just gaps) are thought to exist \citep{Neronov, Levinson}. These compact regions exhibit a large voltage-drop \citep{Hirotani, Wald} and can thus accelerate charged particles to ultra-relativistic energies accompanied by HEP-emission \citep{Ptitsyna}. Recently, it was shown for the case of the peculiar radio-galaxy IC 310, that such gaps are likely to exist in AGN-magnetospheres \citep{Aleksic}. IC 310 showed extreme flux variations (Fig. 4 in \citet{Aleksic}) with time-scales of about 5 min at TeV-energies. This is evidence of emission-regions of length-scales below the Schwarzschild-radius $r_{\rm S}$ and thus can neither be explained by mini-jet models nor by models with jet-cloud-interaction \citep{Aleksic}.\\
In this proceeding, the physical setting of the considered cascade will be stated in all generality in Sec. \ref{GENERAL SETTING}. The kinetic equation of the cascade and its numerical solution via a python3-code will be elucidated in Sec. \ref{THE KINETIC EQUATION} and \ref{NUMERICAL SOLUTION}, respectively. In Sec. \ref{APPLICATION TO MRK 501} the code will be applied for the case of an electromagnetic cascade initiated by electrons, that have been accelerated in a magnetospheric vacuum gap in Mrk 501 and interact with Lyman-$\alpha$-photons.

\section{GENERAL SETTING} \label{GENERAL SETTING}

The electromagnetic cascade considered here have been formulated by Z88. More specific, linear IC-Klein-Nishina-pair-cascades are considered. This means the repeated interaction of relativistic electrons and positrons (henceforth just called electrons) as well as HEPs with an SBP-field via IC-scattering and pair-production (PP). The energy divided by the electron's rest-frame-energy $m_{\rm e} c^2$ for the electrons, the HEPs and the SBPs is denoted by $\gamma$ (Lorentz-factor), $x_\gamma$ and $x$, respectively. Hence, there are three pools of species:
\begin{itemize}
\item The pool of electrons with spectral number-density $N(\gamma)$.
\item The pool of HEPs with spectral number-density $n_\gamma(x_\gamma)$.
\item The pool of SBPs with spectral number-density $n_0(x)$. These SBPs act as target-photons in what follows.
\end{itemize}
These distributions are assumed to be isotropic, homogeneous and time-independent. As interaction-processes, there is solely PP and IC-scattering assumed. (Curvature-radiation, elastic photon-photon-scattering etc. is not included.)\\
PP is assumed to happen solely via collisions of the HEPs with the SBPs (HEP-HEP-interaction, triplet-PP or higher-order PP-processes are neglected.). By this, new electrons are created and supplied to the electron-pool. The electrons, that have been created, are then again available for IC-scattering.\\
IC-scattering happens via collisions of electrons with the SBPs. By this,
\begin{itemize}
\item HEPs are produced and delivered to the pool of HEPs. So, these HEPs are again available for PP.
\item the electrons are down-scattered, i.e. they lose energy, although they remain in the electron-pool and can again IC-scatter.
\end{itemize}
Electron-electron-interactions as well as electron-escape and photon-escape from the reaction-volume is neglected.\\
Via the interplay of PP and IC-scattering a linear cascade develops \citep{NoteLinearCascade}.\\
The cascade affects/replenishes the electron-pool and the pool of HEPs. The pool of SBPs is not affected by the cascade, hence $n_0(x)$ is assumed to be untouched by IC-scattering ($\mathrel{\hat=}$ linearity of the cascade).\\
Additionally, it is assumed that
\begin{enumerate}[A.]
	\item electrons with energy $\gamma \gg 1$ are steadily injected with spectral injection-rate $\dot N_{\rm i}(\gamma)$ (Spectral injection-rate means number-density of injected electrons with energy $\gamma$ per time-interval and per $\gamma$-interval.) and/or that
	\item HEPs with energy $x_\gamma \gg 1$ are steadily injected with spectral injection-rate $\dot n_{\rm i}(x_\gamma)$ (Spectral injection-rate means number-density of injected HEPs with energy $x_\gamma$ per time-interval and per $x_\gamma$-interval.).
\end{enumerate}
into the respective pools. For the injected electrons $\gamma \cdot x > 1$ is assumed to hold, which means that IC-scattering takes place in the Klein-Nishina-regime. This is the reason why the considered type of cascade is called IC-\textsl{Klein-Nishina}-pair-cascade by Z88. Notice however, that repeated IC-scattering lowers the energy of an electron and if the electron's energy has shrunk below the Klein-Nishina-regime, IC-scattering still goes on and on. Therefore, the Klein-Nishina-criterion is not fulfilled for all IC-scattering-events, only for those involving the injected electrons. Furthermore, $x_\gamma \cdot x > 1$ is assumed to be satisfied for the injected HEPs, meaning that the injected photons' energy is above the PP-threshold. The process of such a cascade is sketched in Fig. 1.

\begin{figure}[htbp]
 \includegraphics[width=0.5\textwidth]{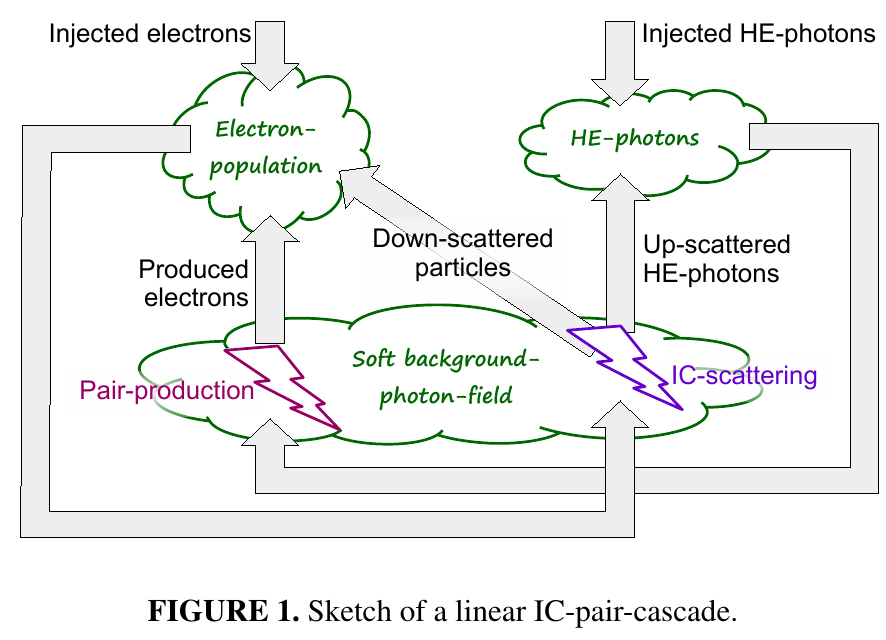}
 \includegraphics[width=0.5\textwidth]{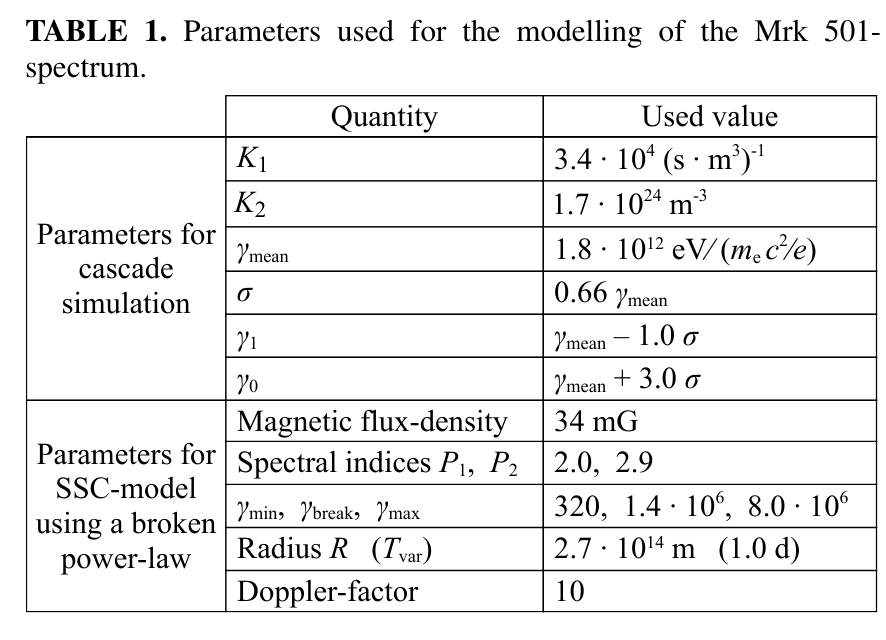}
\end{figure}
\setcounter{figure}{1}

\section{THE KINETIC EQUATION} \label{THE KINETIC EQUATION}

The electrons' kinetic equation of the cascade described in Sec. \ref{GENERAL SETTING} was given by Z88 and can be derived as follows:\\
First, to be able to handle IC-scattering and PP one has to yield spectral interaction-rates, $C$ for IC-scattering and $P$ for PP, respectively. This has been done by Z88 based on findings of \citet{Jones} as well as of \citet{AAN}. In those derivations of $C$ and $P$ isotropy, $\gamma \gg 1$, $x_\gamma \gg 1$ and $x \ll 1$ has been assumed.\\
The spectral IC-interaction-rate $C(\gamma,\gamma')$, i.e. the probability per time-interval and per $\gamma'$-interval, for IC-scattering of an electron with original energy $\gamma$ to final energy $\gamma'$ on the SBP-field $n_0$, is given by Eq. A1 of Z88. (In this equation $E_\gamma$ is not a quantity on its own but a typing error. It means $E \cdot \gamma.$) $C(\gamma,\gamma')$ is an integral along $x$ over $n_0(x)$ times a certain function of $\gamma$, $\gamma'$ and $x$. Hence, as soon as $n_0$ is given, $C$ is given, too. (However, $n_0$ has to be integrable.) Now, as conservation of energy approximately gives $\gamma \approx \gamma' + x_\gamma$ it is obvious that $C(\gamma,\gamma-x_\gamma)$ is the probability per time-interval and per $x_\gamma$-interval, for IC-scattering of an electron with original energy $\gamma$ on the SBP-field with the production of an HEP of energy $x_\gamma$. Notice that $C(\gamma,\gamma-x_\gamma)$ is a function of $\gamma$ and $x_\gamma$, while $C(\gamma,\gamma')$ is dependent on $\gamma$ and $\gamma'$.\\
Similarly, $P(x_\gamma,\gamma)$ is the spectral PP-interaction-rate, i.e. the probability per time-interval and per $\gamma$-interval for production of an electron with energy $\gamma$ due to interaction of an HEP with energy $x_\gamma$ with the SBP-field. Of course, at the same time an electron with energy $\approx x_\gamma - \gamma$ is created, too. $P(x_\gamma,\gamma)$ is given by Eq. B1 of Z88 and is again technically an integral along $x$ over $n_0(x)$ times a certain function of $x_\gamma$, $\gamma$ and $x$. Hence, as soon as $n_0$ is given, $P$ is given, too.\\
Second, with aid of the spectral interaction-rates one can now determine the sources and sinks of the electron-distribution. To be more precise, one can determine the summands of the rate-of-change $\dot N(\gamma)$ of the spectral number-density $N$ of the electrons:
\begin{itemize}
\item Consider electrons with energy $\gamma$. They lose energy via IC-scattering, which leads to a sink of $\dot N$ at energy $\gamma$. Therefore, $\int ^{\gamma}_{1} N(\gamma) C(\gamma,\gamma') \rm{d}\gamma'$ gives the number-density of electrons with any energy below $\gamma$, that are produced via IC-down-scattering of electrons with energy $\gamma$ and with spectral number-density $N$ on the SBP-field, per time-interval and per $\gamma$-interval. Hence, it is the spectral decrease-rate of the number-density of electrons at energy $\gamma$ via IC-down-scattering to lower energies.
\item Now, consider electrons with any energy $\gamma'$ above $\gamma$. They can be IC-down-scattered to the energy $\gamma$, which leads to a source of $\dot N$ at energy $\gamma$. Therefore, $\int _{\gamma}^{\infty} N(\gamma') C(\gamma',\gamma) \rm{d}\gamma'$ gives the number-density of electrons with energy $\gamma$, that are produced via IC-down-scattering of electrons with any energy above $\gamma$ and with spectral number-density $N$ on the SBP-field, per time-interval and per $\gamma$-interval. Thus, it is the spectral increase-rate of the number-density of electrons at energy $\gamma$ via IC-down-scattering from higher energies.
\item Now, consider HEPs with any energy $x_\gamma$ above the PP-threshold $x_{\gamma,\rm{thre}}(\gamma,x_0)$ \citep{NotePPThreshold}. Such HEPs can interact with the SBPs and pair-produce an electron with energy $\gamma$, which causes a source of $\dot N$ at energy $\gamma$. Therefore, $\int _{x_{\gamma,\rm{thre}}(\gamma,x_0)}^{\infty} n_\gamma(x_\gamma) P(x_\gamma,\gamma) \rm{d}x_\gamma$ gives the number-density of electrons with energy $\gamma$, that are pair-produced via interaction of all available HEPs above the PP-threshold with the SBP-field, per time-interval and per $\gamma$-interval. So, it is the spectral increase-rate of the number-density of electrons at energy $\gamma$ via PP.
\item Of course, the injection $\dot N_{\rm i}(\gamma)$ is a source-term, too.
\end{itemize}
Adding up the terms with taking account for the signs yields the spectral rate-of-change of the number-density of the electrons:
\begin{equation}
\dot N(\gamma) = \dot N_{\mathrm{i}}(\gamma) - N(\gamma) \int ^{\gamma}_{1} C(\gamma,\gamma') \mathrm{d}\gamma' + \int _{\gamma}^{\infty} N(\gamma') C(\gamma',\gamma) \mathrm{d}\gamma' + \int _{x_{\gamma,\mathrm{thre}}(\gamma,x_0)}^{\infty} n_\gamma(x_\gamma) P(x_\gamma,\gamma) \mathrm{d}x_\gamma
\label{RateOfChangeElectrons}
\end{equation}
Third, one can determine the sources and sinks of the HEP-distribution. To be more precise, one can determine the summands of the rate-of-change $\dot n_\gamma(x_\gamma)$ of the spectral number-density $n_\gamma$ of the HEPs:
\begin{itemize}
\item Consider HEPs with energy $x_\gamma$. They can pair-produce, which destroys them and thus results in a sink of $\dot n_\gamma$ at energy $x_\gamma$. Therefore, $\int ^{\gamma_{\rm{max}}(x_\gamma,x_0)}_{\gamma_{\rm{min}}(x_\gamma,x_0)} n_\gamma(x_\gamma) P(x_\gamma,\gamma) \rm{d}\gamma$ \citep{NotePPLimits} gives the number-density of HEPs with energy $x_\gamma$, that pair-produce electrons on the SBP-field, per time-interval and per $x_\gamma$-interval. Hence, it is the spectral decrease-rate of the number-density of HEPs at energy $x_\gamma$ via PP.
\item Next, consider electrons with any energy $\gamma$ above the threshold $\gamma_{\rm{thre}}(x_\gamma,x_0)$ for IC-scattering \citep{NoteICThreshold}. They can be IC-down-scattered to the energy $\gamma - x_\gamma$, while IC-up-scattering an SBP to the energy $x_\gamma$. Hence, this is a source of $\dot n_\gamma$ at energy $x_\gamma$. Therefore, $\int _{\gamma_{\rm{thre}}(x_\gamma,x_0)}^{\infty} N(\gamma) C(\gamma,\gamma-x_\gamma) \rm{d}\gamma$ yields the number-density of HEPs with energy $x_\gamma$, that are produced via IC-down-scattering of electrons with any energy above $\gamma_{\rm{thre}}(x_\gamma,x_0)$ and with spectral number-density $N$ on the SBP-field, per time-interval and per $x_\gamma$-interval. Therefore, it is the spectral increase-rate of the number-density of HEPs at energy $x_\gamma$ via IC-up-scattering of SBPs.
\item Of course, the injection $\dot n_{\rm i}(x_\gamma)$ is a source-term, too.
\end{itemize}
Adding up the terms with taking account for the signs yields the spectral rate-of-change of the number-density of the HEPs:
\begin{equation}
\dot n_\gamma(x_\gamma) = \dot n_{\mathrm{i}}(x_\gamma) - n_\gamma(x_\gamma) \int ^{\gamma_{\mathrm{max}}(x_\gamma,x_0)}_{\gamma_{\mathrm{min}}(x_\gamma,x_0)} P(x_\gamma,\gamma) \mathrm{d}\gamma + \int _{\gamma_{\mathrm{thre}}(x_\gamma,x_0)}^{\infty} N(\gamma) C(\gamma,\gamma-x_\gamma) \mathrm{d}\gamma
\label{RateOfChangePhotons}
\end{equation}
In the steady-state, the electron- and the HEP-distribution does not change. Hence, the total rates-of-change of Eq. \ref{RateOfChangeElectrons} and \ref{RateOfChangePhotons} can be set equal to zero. Doing so, solving Eq. \ref{RateOfChangeElectrons} for $N(\gamma)$, solving Eq. \ref{RateOfChangePhotons} for $n_\gamma(x_\gamma)$ and substituting into Eq. \ref{RateOfChangeElectrons} yields the final kinetic equation for the electron-distribution:
\begin{equation}
N(\gamma) = \frac{\dot N_{\mathrm{i}}(\gamma) + \int _{\gamma}^{\infty} N(\gamma') C(\gamma',\gamma) \mathrm{d}\gamma' + \int _{x_{\gamma,\mathrm{thre}}(\gamma,x_0)}^{\infty} \left( \dot n_{\mathrm{i}}(x_\gamma) + \int _{\gamma_{\mathrm{thre}}(x_\gamma,x_0)}^{\infty} N(\gamma) C(\gamma,\gamma-x_\gamma) \mathrm{d}\gamma \right) \frac{P(x_\gamma,\gamma)}{\int ^{\gamma_{\mathrm{max}}(x_\gamma,x_0)}_{\gamma_{\mathrm{min}}(x_\gamma,x_0)} P(x_\gamma,\gamma) \mathrm{d}\gamma} \mathrm{d}x_\gamma}{\int ^{\gamma}_{1} C(\gamma,\gamma') \mathrm{d}\gamma'}
\label{KineticEquation}
\end{equation}
This equation is Eq. 1 of Z88. The fraction in the numerator is equivalent to the quantity $p(x_\gamma,\gamma)$ of Z88.


\section{NUMERICAL SOLUTION} \label{NUMERICAL SOLUTION}

The goal is now to find solutions $N(\gamma)$ to Eq. \ref{KineticEquation}. As noted above, $C$ and $P$ are completely specified as soon as $n_0$ is given. Then, if $n_0$, $\dot N_{\rm i}$ and $\dot n_{\rm i}$ are prescribed, the right-hand side is a functional of solely $N$, or briefly written $N(\gamma) = \mathcal{F}(n_0, \dot N_{\rm i}, \dot n_{\rm i}, N, \gamma)$.\\
The problem can be solved iteratively. Prescribing $n_0$, $\dot N_{\rm i}$ and $\dot n_{\rm i}$ based on the physical setting and guessing an initial function $N_{\rm{init}}(\gamma)$ one can determine the right-hand side $\mathcal{F}$, which is equal to the new $N(\gamma)$, called $N_{0}(\gamma)$. This can again be plugged into $\mathcal{F}$, etc. Hence, $N_{0}(\gamma) = \mathcal{F}(n_0, \dot N_{\rm i}, \dot n_{\rm i}, N_{\rm{init}}, \gamma)$ is the initialisation and $N_{j}(\gamma) = \mathcal{F}(n_0, \dot N_{\rm i}, \dot n_{\rm i}, N_{j-1}, \gamma)$ for $j \in \mathbb{N}$ is the $j^{\rm th}$ iteration step. Employing the Banach fixed-point theorem, it can be shown that convergence is achieved.\\
Next, to yield $N(\gamma)$ in the Thomson-regime (i.e. for $\gamma \ll 1/(4 x_0)$), where no electrons are pair-produced (Z88) and where energy-changes due to IC-scatterings are tiny, Eq. 8 by Z88 is used, which essentially is a separated and integrated continuity-equation. The solution of $N$ around $\gamma \approx 1/(4 x_0)$, where neither the pure Thomson-limit nor the Klein-Nishina-limit is valid, is then determined via interpolating between the $N$ from the continuity-equation with the iterated $N$.\\
We have developed an algorithm using the python3-language that executes the described procedure. As no general analytical expressions for the integrals are known, they have to be computed numerically. To be able to determine the integrals with infinite integration-range in a finite timespan, one has to restrict the infinite integration-ranges to finite ranges. This is possible without doing errors, if the functions $n_0$, $\dot N_{\rm i}$ and $\dot n_{\rm i}$ are not infinitely extended but have an upper cut-off beyond which the distributions are vanishing. Hence, as input to the code one has to prescribe three distributions in energy-space:
\begin{itemize}
\item That one $\dot N_{\rm i}(\gamma)$ of the injected electrons.
\item That one $\dot n_{\rm i}(x_\gamma)$ of the injected HEPs.
\item That one $n_0(x)$ of the SBPs.
\end{itemize}
Every kind of integrable function can be processed by the code. 
\\
Then, the code determines the steady-state electron-distribution $N(\gamma)$ that evolves. Additionally, the distribution of produced HEPs (second summand in the parentheses of Eq. \ref{KineticEquation}) is computed. Certainly, despite of the assumption of no HEP-escape (cf. Sec. \ref{GENERAL SETTING}), HEPs can escape from the boundary-shells of the interaction-volume. These photons can then be observable.

\section{APPLICATION TO MRK 501} \label{APPLICATION TO MRK 501}

\begin{figure}
 \includegraphics[width=1.0\textwidth]{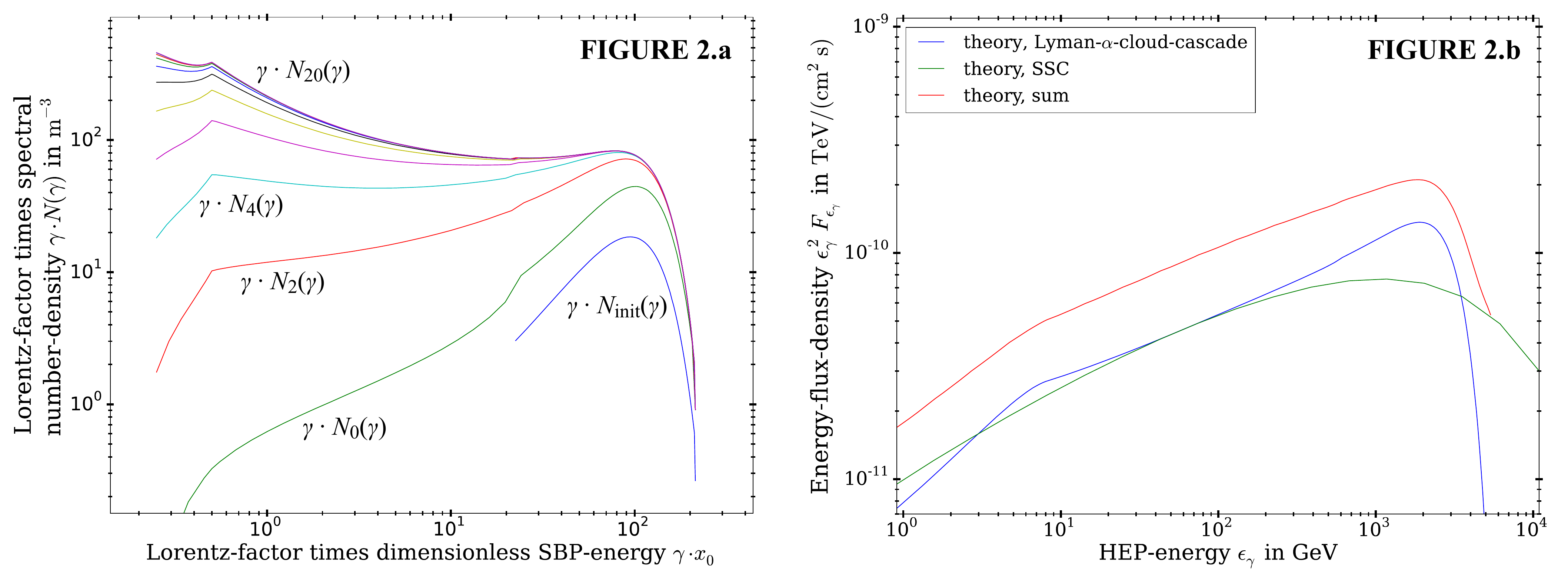}
 \caption{\textbf{a:} Exemplary plot of the electron-distribution versus electron dimensionless energy in Klein-Nishina-regime. The smaller $\gamma \cdot x_0$ (i.e. the nearer to the Thomson-regime), the more iterations are necessary, which is the case as scatterings cause only smaller energy-transfers. \textbf{b:} Exemplary plot of the modelled energy-flux-density versus HEP-energy.}
 \label{CascadeIterationPlusSpectrum}
\end{figure}

The general scenario sketched in Sec. \ref{GENERAL SETTING} will now be specified for the following setting. A presumed gap in the magnetosphere of Mrk 501 accelerates electrons (gaussianly distributed in the energy-space), that leak into it from the accretion disk, to energies of about $\gamma_{\rm{mean}}$. Hence, they have a distribution 
\begin{equation}
\dot N_{\rm i}(\gamma) = \left\{
\begin{array}{ll}
\frac{K_1}{\sigma \sqrt(2 \pi)} \cdot \exp \left( -\frac{(\gamma-\gamma_{\mathrm{mean}})^2}{2 \sigma^2} \right) & \mathrm{if} \; \gamma_1 \leq \gamma \leq \gamma_0 \mathrm{,} \\
0 & \mathrm{otherwise.}
\end{array}
\right.
\label{DistributionElectrons}
\end{equation}
where $\gamma_1$ and $\gamma_0$ is the lower and upper cut-off, respectively, satisfying $1 \ll \gamma_1 < \gamma_0$ in accordance with the condition of bullet point "A" in Sec. \ref{GENERAL SETTING}. $\sigma$ is the width of the distribution and $K_1$ is its norm and specifies the total number-density of electrons that are injected per time-interval. These electrons travel away from the AGN-centre along the jet. Furthermore it is assumed that these electrons penetrate into a region, which is pervaded by Lyman-$\alpha$-photons, i.e. by photons of wavelength $121.5 \, \rm{nm}$. Lyman-$\alpha$-photons are abundant in galaxy-cores. Certainly, ionising radiation from the accretion disk or from hot stars affects infalling hydrogen causing recombination lines. The kink in the radio jet of Mrk 501 and its fanning-out \citep{Conway} might be evidence for a violent event (galaxy-merger in the past, binary black hole or jet-star-interaction), so young stars and/or emission nebulae should be present, which justifies the assumption of Lyman-$\alpha$-photons. Thus, the SBPs are distributed according to  
\begin{equation}
n_{0}(x) = K_2 \cdot \delta_{\mathrm{Dirac}} \left( x - x_0 \right)
\label{DistributionSoftPhotons}
\end{equation}
with $x_0 = h/(121.5 \, {\rm{nm}} \, m_{\rm{e}} c)$. Here, $K_2$ describes the total number-density of SBPs and is treated as a free parameter, in contrast to Z88, who uses $K_2 = K_1/(c \cdot \sigma_{\rm Thom})$. In this scenario, no HEPs are assumed to be injected, so $\dot n_{\rm i}(x_\gamma) = 0$. In the interaction-region, the described cascade takes place and HEPs escape from its border-region. We have simulated this cascade with our code. As can be seen in Fig. \ref{CascadeIterationPlusSpectrum}.a the electron-distribution converges after approximately 20 iteration steps and steady-state spectra evolve. The escaping cascaded HEPs as well as photons produced in a traditional synchrotron-self-Compton- (SSC-) scenario taking place farther outwards in the jet form an observable spectrum. The superposition of the cascaded spectrum \citep{NoteCascadedSpectrum} with the SSC-spectrum is shown in Fig. \ref{CascadeIterationPlusSpectrum}.b. The SSC-spectrum constitutes the background from X-rays over $\gamma$-rays up to the TeV-regime, while the cascaded spectrum causes a peculiar bump between 800 GeV and 4 TeV. Such a narrow feature would be difficult to explain with conventional theoretical scenarios, hence its detection would argue for the taking place of the described cascaded. In Tab. 1 the parameters used for the modelling are given. The underlying code used for SSC-modelling is described in \citep{SSCref}.

\section{ACKNOWLEDGEMENT}
The research and writing of this work was partially funded by the project "Promotion inklusive" of the \textit{Universit\"{a}t zu K\"{o}ln} and the German \textit{Bundesministerium f\"{u}r Arbeit und Soziales}. C. W. is grateful for this support. This work utilises matplotlib \citep{matplotlib}. The authors acknowledge all contributions to the matplotlib-project.


\begin{thebibliography}{99}
\bibitem[]{NoteLinearCascade}A cascade is called linear, if self-interaction of cascade-species is negligible, i.e. if solely interactions of electrons or HEPs with the SBPs take place (but not interactions of electrons or HEPs with HEPs). In other words, a cascade is linear if only the SBPs act as target-photons. In this case the production-rate of HEPs, of electrons is direct proportional to the injection-rate of electrons, of HEPs, respectively, which clarifies the denotation "linear". One can show by comparison of cross-sections, that a cascade in the Klein-Nishina-regime, Thomson-regime is usually linear, always non-linear, respectively (Z88). A linear cascade is present, if the produced HEPs' density is similar or smaller in comparison to the SBPs' one. Linear cascades belong to type i-cascades in the classification of \citet{Svensson}.
\bibitem[]{NotePPThreshold}Consider the PP of an electron with energy $\gamma$ from an HEP of energy $x_\gamma$ and an SBP with energy $x$. This is possible only if $x_\gamma \geq x_{\gamma,\rm{thre}}(\gamma,x) := \gamma/(1 - (1/(4 \gamma x)))$ due to kinematics of the PP-event (Z88). Hence, if an extended distribution of SBPs is considered, with $x_0$ being the upper cut-off of this distribution, then PP sets in if $x_\gamma \geq x_{\gamma,\rm{thre}}(\gamma,x_0)$.
\bibitem[]{NotePPLimits}Consider PP by collision of an HEP with energy $x_\gamma$ with an SBP of energy $x$. Such events can produce only such electrons whose energy $\gamma$ satisfies $ x_\gamma(1 - (1 - 1 /(x_\gamma x))^{1/2})/2 =: \gamma_{\rm{min}}(x_\gamma,x) < \gamma < \gamma_{\rm{max}}(x_\gamma,x) := x_\gamma(1 + (1 - 1 /(x_\gamma x))^{1/2})/2$, due to kinematic constraints (Z88). This was concealed in Eq. B11 of Z88. Hence, if an extended distribution of SBPs is considered, with $x_0$ being the upper cut-off of this distribution, then electrons can attain energies $\gamma$ in the maximally extended range $\gamma_{\rm{min}}(x_\gamma,x_0) < \gamma < \gamma_{\rm{max}}(x_\gamma,x_0)$.
\bibitem[]{NoteICThreshold}Imagine an electron with energy $\gamma$, that IC-up-scatters an SBP of energy $x$ to energy $x_\gamma$. For kinematical reasons, this is possible only if $\gamma \geq \gamma_{\rm{thre}}(x_\gamma,x) := x_\gamma(1 + (1 + 1 /(x_\gamma x))^{1/2})/2$. Hence, if the SBP-field is extended in energy-space, with $x_0$ being the maximum energy in this distribution, then IC-scattering sets in for $\gamma \geq \gamma_{\rm{thre}}(x_\gamma,x_0)$.
\bibitem[]{NoteCascadedSpectrum}Essentially, the second summand in the parentheses of Eq. \ref{KineticEquation} is converted to a spectral flux-density, where $2 \cdot 10^9 M_{\odot}$, $r_{\rm S}^3$, $1 \rm{sr}$ and $4 \cdot 10^8 \rm{ly}$ is used for the mass of the central supermassive black hole of Mrk 501, for the emitting volume, for the solid angle, into which the radiation is focused, and for the distance from Mrk 501 to Earth, respectively.
\end{thebibliography}
\end{document}